\documentclass[aps,prl,twocolumn,amsmath,amsart,amssymb,superscriptaddress]{revtex4}
\usepackage{graphicx}
\usepackage[T1]{fontenc}
\usepackage{amsmath}
\usepackage{color}
\usepackage{times}

\usepackage{ulem}

\newcommand{\BFCA}{Ba(Fe$_{1-x}$Co$_x)_2$As$_2$}
\newcommand{\BFA}{BaFe$_2$As$_2$}
\newcommand{\FSS}{FeSe$_{1-x}$S$_{x}$}

\newcommand{\BFAP}{BaFe$_2$(As$_{1-x}$P$_x$)$_2$}

\begin{document}


\title{Evolution of pairing symmetry in {\FSS} as probed by uniaxial-strain tuning of $T_c$}


\author{Ruixian Liu\textsuperscript{\S}}
\author{Qi Tang\textsuperscript{\S}}
\author{Chang Liu}
\author{Chunyi Li}
\author{Kaijuan Zhou}
\author{Qiaoyu Wang}
\author{Xingye Lu}
\email{luxy@bnu.edu.cn}
\affiliation{Center for Advanced Quantum Studies, School of Physics and Astronomy, Beijing Normal University, Beijing, 100875, China}

\date{\today}

\begin{abstract}

In iron-based superconductors (FeSCs), the interplay between electronic nematicity and superconductivity is essential for understanding the exotic superconducting ground state.
In the nematic regime, uniaxial-strain ($\varepsilon$) tuning of the superconducting transition temperature $T_c$ [$\Delta T_c(\varepsilon)=\alpha\varepsilon+\beta\varepsilon^2$] offers a unique approach to investigating the evolution of pairing symmetry if both $s$ and $d$ wave pairing instabilities are relevant.
Here, we employ uniaxial strain to tune the $T_c$ of {\FSS}, in which both nematicity and superconductivity undergo significant changes with doping.
While $T_c$ is usually suppressed quadratically with $\varepsilon$ in optimally doped {\BFA}, $\Delta T_c(\varepsilon)$ in {\FSS} dominated by $\Delta T_c(\varepsilon)=\beta\varepsilon^2$ changes its sign from $\beta$ < $0$ in FeSe to $\beta$ > $0$  in {\FSS} ($x\gtrsim0.10$), indicating an evolution of the pairing symmetry from an $s_{\pm}$ state towards an $s+d$ wave state. These findings highlight the $\Delta T_c(\varepsilon)$ as a powerful probe for elucidating the superconducting pairing symmetry in the nematic regime of FeSCs and provide new insights into the evolution of pairing symmetry in FeSCs.

\end{abstract}

\maketitle

The electronic nematic state is one of the most important emergent orders in iron-based superconductors (FeSCs) \cite{Fernandes2014, Bohmer2022, Fernandes2022}. While the origin of this state remains debated, nematic fluctuations are considered a potential pairing mechanism for superconductivity, and electronic nematicity is thought to significantly influence the superconducting properties. Therefore, understanding the interplay between electronic nematicity and superconductivity is crucial for unraveling the mystery of high-temperature ($T_c$) superconductivity in these materials \cite{Fernandes2022}.

{\FSS} is a unique system where magnetic order is absent, allowing nematic order and fluctuations to interact directly with superconductivity \cite{Fernandes2022, Bohmer2018, Coldea2018, Coldea2021}. Figure 1(a) illustrates the phase diagram of {\FSS}. The parent compound, FeSe, undergoes a tetragonal-to-orthorhombic structural transition at $T_s \approx 90$ K, below which a $C_2$-symmetric electronic nematic phase emerges, with the nematic transition temperature $T_{\rm nem} = T_s$. In this nematic phase, FeSe enters a superconducting ground state at $T_c \approx 8$ K \cite{Hsu2008, McQueen2009}. Upon sulfur doping, the orthorhombicity and nematic order are gradually suppressed and vanish around $x \approx 0.17$. Beyond this regime, nematic fluctuations persist over a broad range of temperatures and doping levels \cite{Hosoi2016, Liu2024}, becoming especially pronounced near $x \approx 0.17$, a concentration identified as the nematic quantum critical point (NQCP) \cite{Hosoi2016, Licciardello2019}.

The Fermi surface of {\FSS} consists of two nested hole pockets at the $\Gamma$ point, along with electron pockets at the X and Y points \cite{Coldea2021}. Fermi surface nesting between electron pockets and central hole pockets ($\Gamma$-X and $\Gamma$-Y) generates stripe-type spin fluctuations at $(\pi, 0)/(0, \pi)$ \cite{Dai_RMP},  which potentially lead to an $s_{\pm}$-wave pairing symmetry. Additionally, the N\'eel-type spin fluctuations at $(\pi,\pi)$, likely arising from particle-hole scattering between the electron pockets located at the X and Y points (X-Y) \cite{Chen19, Lu2022, Wang2016, Liu2024_2}, suggest the possibility of $d$-wave pairing symmetry \cite{Fernandes2018}. Scanning tunneling microscopy (STM) and angle-resolved photoemission spectroscopy (ARPES) studies on FeSe have revealed highly anisotropic superconducting gaps \cite{Sprau2017, Liu2018, Hashimoto2018}, which can be described by an extended $s+d$ wave gap function \cite{Liu2018}.

In {\FSS}, both thermal conductivity and specific heat measurements have revealed a pronounced change in the superconducting gap structure near the NQCP \cite{Sato2018}. STM investigations have further identified two distinct superconducting pairing states on either side of the NQCP, indicating a significant evolution of the pairing symmetry or gap structure across this critical point \cite{Sato2018, Hanaguri2018, Matsuura2023}. However, despite these findings, the detailed nature of this evolution remains insufficiently explored.

While directly obtaining evidence for the doping evolution of gap symmetry in {\FSS} remains highly challenging, the interplay between electronic nematicity and superconductivity offers an alternative approach to address this issue. Theoretically, Fernandes \textit{et al.} proposed that electronic nematicity could be used to probe superconducting pairing in FeSCs, as a $d$-wave instability may compete with the dominant $s$-wave instability, given that their free energies could be comparable \cite{Fernandes13}. By incorporating a trilinear coupling term, $F_{\rm SC-nem} \propto \varphi \Delta_s \Delta_d \cos\theta$, between nematicity ($\varphi$) and the $s$- and $d$-wave order parameters ($\Delta_s$ and $\Delta_d$) in spin fluctuation Eliashberg calculations, Ref. \cite{Fernandes13} examined the effects of tunable nematicity on $s$- and $d$-wave superconductivity. They predicted that nematicity-induced changes in $T_c$ [$\Delta T_c(\varphi)$] could serve as a probe to distinguish between $s$- and $d$-wave pairing regimes.

Multicomponent superconductivity, involving a mixture of $s$- and $d$-wave pairing symmetries, has recently been observed in {\BFCA} through elastocaloric measurements under uniaxial strain, confirming the competition between $s$- and $d$-wave pairing in iron-pnictide superconductors \cite{Fisher2024}. A subsequent theoretical study by Fernandes \textit{et al.} suggested that similar competition is expected to occur in FeSe, leading to an evolution of pairing symmetry driven by specific tuning parameters. This is consistent with the anisotropic superconducting gap and the doping-dependent evolution of the gap structure in {\FSS} \cite{Liu2018, Sato2018}.

\begin{figure}[htbp!]
\includegraphics[width=8cm]{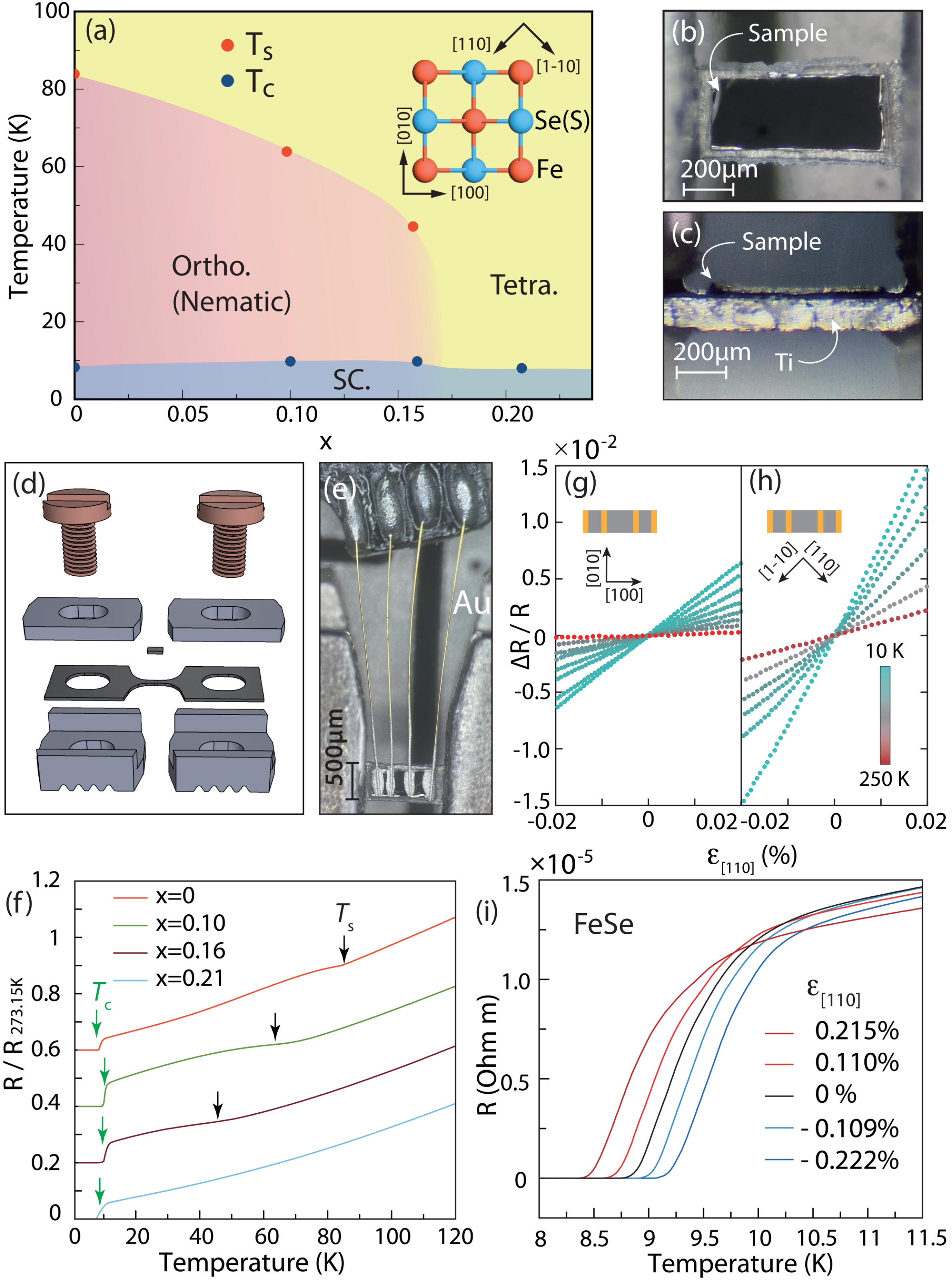}
\caption{
(a) Phase diagram of {\FSS}. The horizontal axis represents sulfur doping concentration. Orange and blue dots indicate the structural transition $T_s$ and superconducting transition temperature $T_c$, respectively. The inset shows a schematic of the lattice structure in a two-iron unit cell. (b)-(c) Photographs of a thin {\FSS} single crystal affixed on a titanium bridge. (d) The sandwich confirguration fixing the titanium platform. (e)  Four-electrode method for measuring the resistivity of strained {\FSS} crystals. (f) Normalized resistance-temperature (RT) curves for {\FSS} with doping concentrations $x$ = 0, 0.10, 0.16, and 0.21. Black and green arrows mark the $T_s$ and $T_c$, respectively. (g), (h) Strain-dependent $\Delta R/R$ measured under strains applied along the $[100]$ or $[110]$ directions for an $x=0.10$ sample. Different colors correspond to different temperatures, as indicated by the colorbar. (g) $\varepsilon_{[110]}$ tuning of resistance and $T_c$ in FeSe. }
\label{fig1}
\end{figure}


%
It is well established that uniaxial strain along the tetragonal [110] direction ($\varepsilon_{[110]}$) couples directly to the nematic order parameter  $\varphi$, making it an effective tuning parameter for electronic nematicity. In this study, we utilize precisely and continuously tunable uniaxial strain as a tuning parameter for $T_c$ in {\FSS}. The functional form of $\Delta T_c(\varphi)$ [equivalent to $\Delta T_c(\varepsilon)$], dictated by the coupling $F_{\rm SC-nem}\propto \varphi \Delta_s \Delta_d \cos\theta$, allows us to unveil the evolution of underlying pairing symmetry in the nematic regime of {\FSS}.
In undoped and lightly doped {\FSS}, $\varepsilon_{[110]}$-induced change in $T_c$ is dominated by $\Delta T_c(\varepsilon)=\beta\varepsilon^2$ with $\beta<0$, consistent with the predominant $s_{\pm}$ wave pairing. At higher doping, $\beta$ starts to evolve to $\beta>0$ and $\Delta T_c(\varepsilon)$ is well characterized by $\Delta T_c(\varepsilon)=-|\beta|\varepsilon^2$ across the NQCP, indicating a transition from $s_{\pm}$ pairing towards a mixed state of $s_{\pm}$- and $d$-wave pairing instabilities.
Our results confirm the role of electronic nematicity as a probe for pairing symmetry and reveal that the pairing symmetry in {\FSS} evolves because of the competition between $s$- and $d$-wave pairing instability in this system, highlighting the need for further spectroscopic studies under uniaxial strain.

\begin{figure}[htbp!]
\includegraphics[width=8cm]{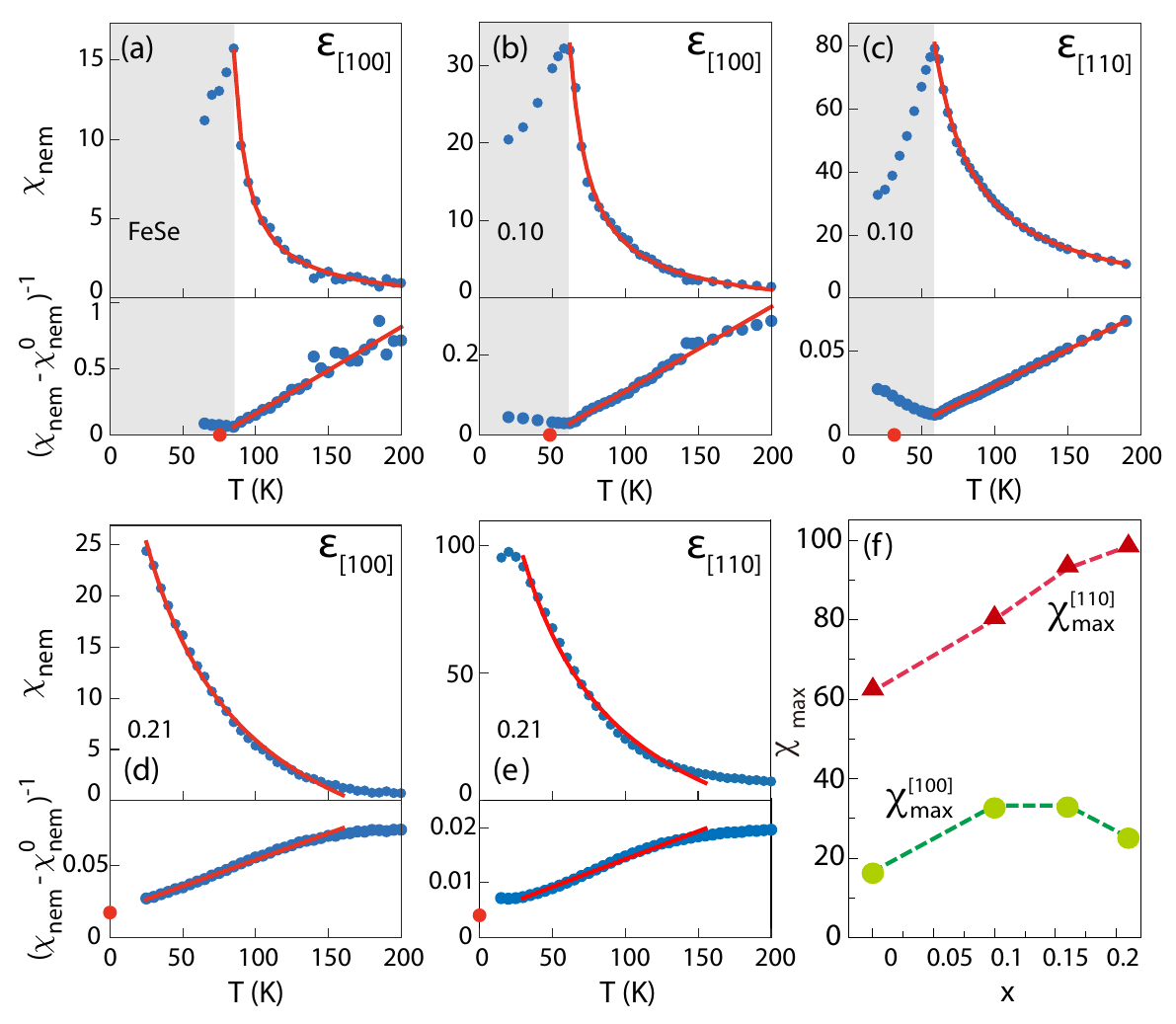}
\caption{
Nematic susceptibility $\chi_{\rm nem}(T)$ extracted from elastoresistance measurements \cite{SI}. The red solid lines are Curie-Weiss fitting $\chi_{\rm nem}(T)=\frac{\lambda}{a(T-T^*)}+\chi^0$. The linear dependence regions in the lower panels show the quality of the fitting. The red dots on the horionzontal axis indicate the fitted $T^*>0$, while the red dots on the vertical axis in (d) and (e) indicate $T^*<0$. (f) Maximum values of nematic susceptibility measured under different strain directions. The red triangles represent strain applied along the [110] direction, while the green dots correspond to strain in the [100] direction. A comparison reveals that the red triangles, corresponding to $B_{2g}$ nematic fluctuations, dominate in the {\FSS} system.}
\label{fig2}
\end{figure}

\begin{figure*}[htbp!]
\includegraphics[width=16 cm]{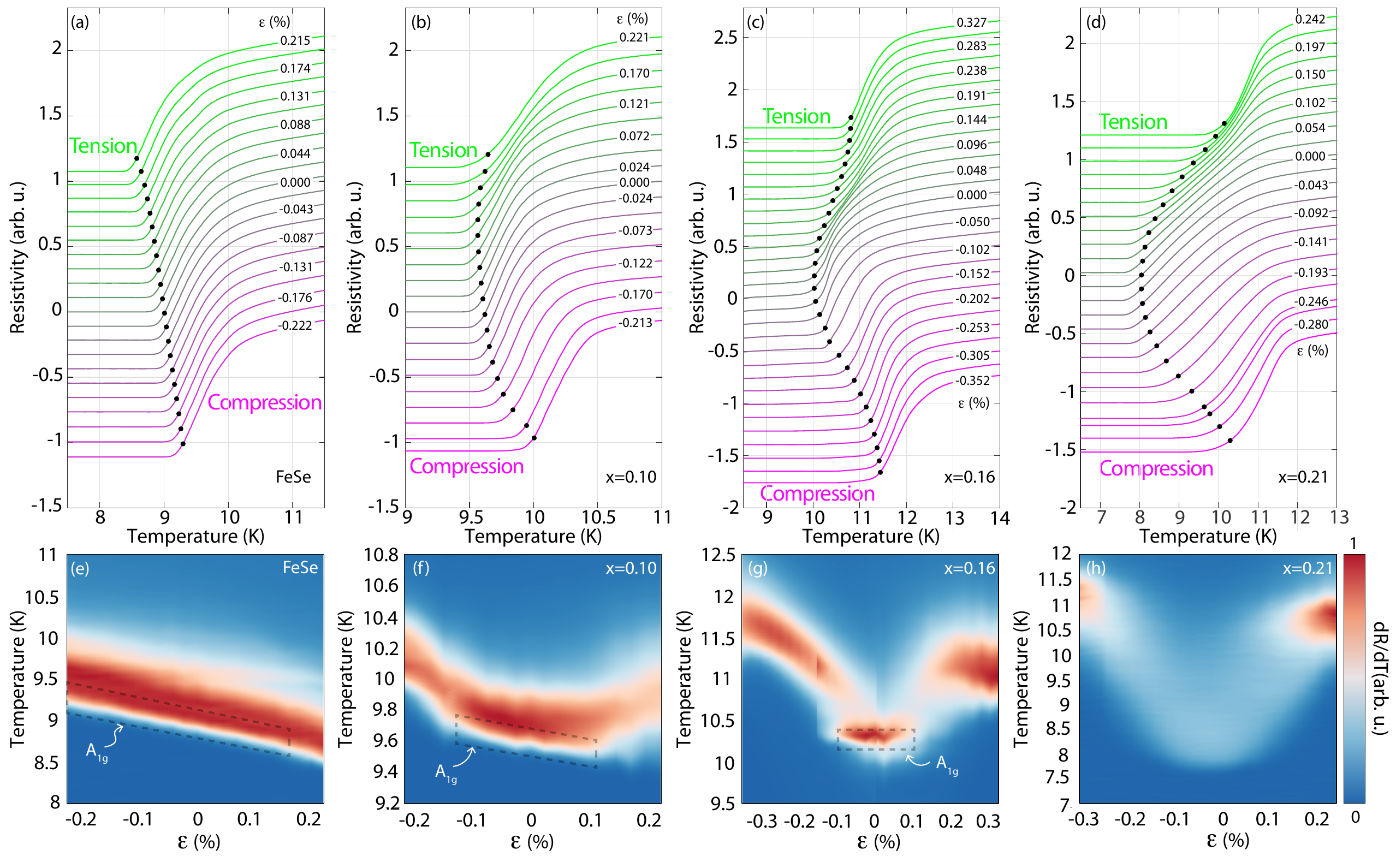}
\caption{
(a)-(d) Normalized resistance-temperature (RT) curves under tensile and compressive strain for samples with different doping concentrations. The zero points of different curves are vertically shifted according to their corresponding strain magnitudes. The color of each curve corresponds to the magnitude of the strain. Black dots indicate the determined superconducting transition temperatures. (e)-(h) Colormap of dR/dT for samples with varying doping concentrations. The areas marked by grey dashed boxes indicate regions obscured by the $A_{1g}$ channel due to twinning, exhibiting linear behavior.}
\label{fig3}
\end{figure*}

The {\FSS} single crystals used in this study were grown using the chemical vapor transport method, as described in detail elsewhere \cite{Liu2024}. The doping levels of the samples were determined by Energy-Dispersive X-ray Spectroscopy measurements. To cover the nematic regime, we selected four doping levels ($x$ = 0, 0.10, 0.16, and 0.21) across the NQCP for this study [Fig. 1(b)]. The corresponding phase transition temperatures are indicated by solid circles in the phase diagram of {\FSS} [Fig. 1(a)].
Uniaxial strain was applied using bowtie-shaped titanium platforms ($0.15$ mm in thickness) as substrates, following the procedure outlined in Ref. \cite{Bartlett2021}. Specifically, thin, flat {\FSS} single crystals (approximately 10 $\mu$m in thickness) were affixed to the neck of the titanium platform using EP29LPSP epoxy, as illustrated in Figs. 1(b) and 1(c) \cite{Yang2023}. The titanium platform was subsequently mounted in a sandwich configuration onto a commercial FC100 strain cell, allowing for the application of either tensile or compressive strain \cite{SI}. Previous studies have demonstrated that strain applied to the titanium platform can be effectively transmitted to the thin crystal affixed to it \cite{Bartlett2021, Yang2023}. In our measurements, the uniaxial strain was calculated from the force applied by the strain cell and further confirmed using the cryogenic digital image correlation method described in Ref. \cite{Mo2024}.
Figure 1(d) presents the normalized resistance-temperature (RT) curves, with the structural phase transition temperature $T_s$ marked by black arrows. The abrupt change in resistance at lower temperatures signifies the onset of superconductivity. Figure 1(f) illustrates the RT curve of a FeSe sample under uniaxial strain $\varepsilon_{[110]}$, demonstrating that in this system, uniaxial strain not only alters $T_{c, {\rm onset}}$ and $T_{c, {\rm offset}}$ but also affects the resistance values in the normal state (above $T=11$ K in this figure). This suggests that relying on a fixed resistance value to determine the shift of $T_c$ may lead to inaccuracies. Therefore, We chose the temperature with 10$\%$ of the resistance at the normal state ($T\approx12$ K)  to track the relative change of $T_c$ under uniaxial strain \cite{Bartlett2021}.

Before investigating the interplay between nematicity and superconductivity, we first validate the doping evolution of the nematic susceptibility within the nematic fluctuating regime in {\FSS}. Hosoi {\it et al.} have previously measured the elastoresistivity of {\FSS}, revealing widespread nematic susceptibilities $\chi_{\rm nem}$ in the $B_{2g}$ channel ($\chi_{\rm nem}^{B_{2g}}$), which is significantly enhanced around the NQCP \cite{Hosoi2016}.
In our measurements, we employed longitudinal elastoresistance measurements to probe the $\chi_{\rm nem}$. The change in resistivity induced by uniaxial strain --- $(\Delta R/R)_{xx}$ ($\varepsilon\parallel I \parallel x$) --- was found to vary linearly within the elastic limit, which enabled us to determine the nematic susceptibility by analyzing the slope of $(\Delta R/R)_{xx}$ versus $\varepsilon_{xx}$ \cite{Chu2012, Kuo2016, SI}. This method has been demonstrated to be effective in probing the divergent nematic susceptibility in {\BFCA} \cite{Chu2012}.


\begin{figure*}[htbp!]
\includegraphics[width=16cm]{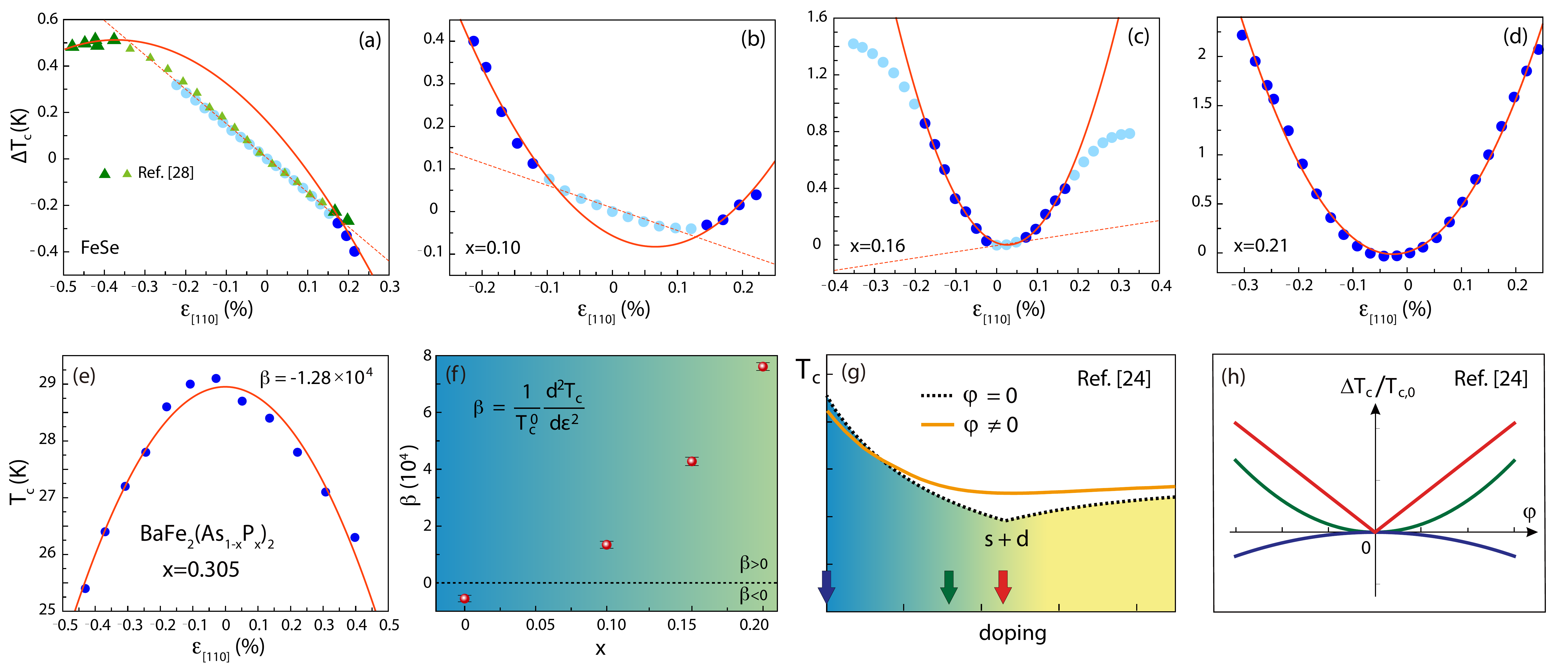}
\caption{
(a)-(d) Quadratic fitting of $\Delta T_c(\varepsilon)$ for {\FSS}, represented by red curves, based on the data points (blue dots) in each panel. Data points excluded from the analysis are that either fall within the twin boundary or exceed the range of infinitesimal strain approximation. In (a), points marked with triangles are taken from Ref.  \cite{Bartlett2021}. (e) Quadratic fitting analysis of the superconducting transition temperature under strain control for BaFe$_2$(As$_{1-x}$P$_x$)$_2$. (f) Doping evolution of the normalized quadratic coefficient $\beta$ in {\FSS}, showing a sign change from negative to positive with increasing sulfur content $x$, followed by a progressive enhancement. The gradation in background color correlates with the interplay between $s$-wave and $d$-wave superconducting states in the system. (g)-(h) Predictions from a multi-band spin fluctuation Eliashberg model considering the nematic order parameter for $T_c$ tuning.
}
\label{fig4}
\end{figure*}

The extracted nematic susceptibilities measured along tetragonal [100] and [110] directions of $x=0, 0.10$, and $0.21$ are displayed in Figs. 2(a) to 2(e). Following the analysis described in Ref. \cite{Chu2012}, we performed Curie-Weiss fitting of the data using $\chi_{\rm nem}(T)=\frac{\lambda}{a(T-T^*)}+\chi^0_{\rm nem}$, where $\lambda$ is a linear coupling constant between nematicity and uniaxial strain, and $T^*$ is the bare nematic transition temperature. The fitting results, represented by red curves, are plotted onto the data points in the upper panels of Figs. 2(a)-2(e). The lower panels present the linear fitting of $\frac{\lambda}{a}(T-T^*)=(\chi_{\rm nem}-\chi^0_{\rm nem})^{-1}$. The intersections between the linear fitting and the temperature axis yield the values of $T^*$, which are positive for $x=0-0.16$ [red dots in Figs. 2(a)-2(c)] and become negative across the NQCP in $x=0.21$ [Figs. 2(d) and 2(e)] \cite{SI}. This doping dependence is consistent with previous reports in FeSe and iron-pnictide superconductors  \cite{Chu2012, Hosoi2016}.

Figure 2(f) summarizes the doping dependence of the nematic susceptibility $\chi_{\rm nem}$, showing the maximum values measured along both the [110] and [100] directions. The susceptibility along the [110] direction,  $\chi_{\rm nem}^{B_{2g}}$, increases monotonically with doping and is constantly much larger than that along the [100] direction,  $\chi_{\rm nem}^{B_{1g}}$ \cite{SI}. These results are in agreement with previous findings reported in Ref. \cite{Hosoi2016}. The persistence of $\chi_{\rm nem}^{B_{2g}}$ provides a strong foundation for investigating the interplay between nematicity and superconductivity.

After illustrating the $\chi_{\rm nem}^{B_{2g}}$ in {\FSS}, we proceed to investigate the uniaxial strain dependence of  $T_c$. Figures 3(a)-3(d) present the resistivity curves for doping levels  $x = 0, 0.10, 0.16$ and $0.21$, measured under varying uniaxial strain $\varepsilon_{[110]}$.
The superconducting transition temperatures,  $T_c$, marked by black solid dots in each figure, exhibit systematic shifts in response to $\varepsilon_{[110]}$. The strain-induced change of $T_c$ --- $\Delta T_c(\varepsilon_{[110]})$, shows  linear strain dependence in $x=0$ while at higher doping levels, it takes on a quadratic-like shape. These behaviors become clearer in the first-order temperature derivative of the resistivity ($dR/dT$) as shown in Figs. 4(e)-4(h).

Uniaxial strain tuning of phase transition temperatures (such as $T_s$ and $T_c$) in the study of FeSCs is a well-established experimental method \cite{Bartlett2021, Ghini21, Fisher2021, Malinowski2020}. To obtain a quantitative understanding, it is necessary to decompose the strain-tuning effects into all the irreducible ($A_{1g}$, $B_{1g}$ and $B_{2g}$) strains under the $D_{4h}$ point group that describes the structure of the unit cell of {\FSS}. Due to the Poisson's effect, the longitudinal strain $\varepsilon_{xx}=\varepsilon_{[110]}$ generates $\varepsilon_{yy}=-\nu\varepsilon_{[110]}$ ($\nu\approx0.32$ for the titanium substrate), resulting in an in-plane symmetry preserving strain $\varepsilon_{A_{1g}}=\frac{1}{2}(\varepsilon_{xx}+\varepsilon_{yy})=\frac{1-\nu}{2}\varepsilon_{xx}=0.34~\varepsilon_{[110]}$ and anti-symmetric strain $\varepsilon_{B_{2g}}=\frac{1}{2}(\varepsilon_{xx}-\varepsilon_{yy})=\frac{1-\nu}{2}\varepsilon_{xx}=0.66~\varepsilon_{[110]}$.
In the small range of $\varepsilon_{[110]}$, it has been demonstrated that the relative change of $T_c$ can be described by a quadratic function $\Delta T_c(\varepsilon) = a \varepsilon_{A_{1g}}+b\varepsilon_{B_{2g}}^2$, where the symmetric $\varepsilon_{A_{1g}}$ contributes only to the linear term, while the antisymmetric $\varepsilon_{B_{2g}}$ dominates the biquadratic term \cite{Malinowski2020, Fisher2021}. In terms of $\varepsilon=\varepsilon_{[110]}$, the formula becomes $\Delta T_c(\varepsilon) = \alpha \varepsilon+\beta\varepsilon^2$, in which $\alpha=0.34 a$ and $\beta = 0.44 b$.

The $dR/dT$ color map Figs. 4(e)-4(h) reveal a pure linear $\Delta T_c(\varepsilon)$ ($\varepsilon = -0.25\%$ to +0.25$\%$) in FeSe [Fig. 4(e)], attributed to $\varepsilon_{A_{1g}}$-induced $\Delta T_c$ during the detwinning process in a prior study of uniaxial-strained FeSe \cite{Bartlett2021}. The strain range corresponding to the linear $\Delta T_c(\varepsilon)$ defines the twin boundary in the phase diagram of $\Delta T_c(\varepsilon)$ \cite{Bartlett2021}. Indeed, such a linear region (associated with twinning) gradually shrinks with increasing (or decreasing) of the doping ($T_s$) and disappears across the NQCP [Fig. 4(h)]. To effectively explore the interplay between nematicity and superconductivity, it is crucial to bypass the linear region and measure the effects of $\varepsilon_{B_{2g}}$ on $T_c$.

Figures 5(a)-5(d) illustrate the strain tuning of $T_c$ extracted from Fig. 4 [black dots in Figs. 4(a)-4(d)]. The data beyond the twin boundary in these figures (blue dots) can be fitted using  $\Delta T_c=\alpha\varepsilon+\beta\varepsilon^2$, where $\varepsilon=\varepsilon_{[110]}$. In Fig 5(c), data points in large strain regions deviated from the quadratic function are excluded. This is due to the requirement of an infinitesimal strain approximation when decomposing $T_c$ tuning results into quadratic functions, following the $D_{4h}$ irreducible representations framework.

The normalized quadratic coefficient $\beta$ derived from our fitting represents the $\varepsilon_{B_{2g}}$ sensitivity of $T_c$. In FeSe, the combined fitting of our data and that from prior study in Ref. \cite{Bartlett2021} generates a negative $\beta$ ($\beta<0$). Note this value might contain large uncertainty due to the large detwinning boundary in FeSe.
Negative quadratic coefficient in $\Delta T_c=\alpha\varepsilon+\beta\varepsilon^2$ have been reported in {\BFCA} \cite{Malinowski2020} and {\BFAP} \cite{Zhao2023}, that is, $T_c$ is suppressed by $\varepsilon_{[110]}$. To facilitate a direct comparison, we measured the $\varepsilon_{[110]}$ tuning of $T_c$ using the same strain cell and obtained $\beta\approx-1.28\times10^4$ following the same analyzing method.
Surprisingly, as the increase of the doping in {\FSS}, the quadratic coefficient $\beta$ evolves to positive values at $x\gtrsim0.1$, indicating that $T_c$ is enhanced by $\varepsilon_{[110]}$ [Fig. 4(e)]. At higher doping, $\beta$ increases with doping, evolves monotonically across the NQCP, and reachs $\beta\approx8\times10^4$. It is worth noting that the linear coefficient $\alpha$ in the optimally doped iron pnictides and in {\FSS} is much smaller, suggesting marginal effects of the symmetric strain $\varepsilon_{A_{1g}}$ in tuning $T_c$ in these systems \cite{ftnote}. 

The $\Delta T_c(\varepsilon)$ in {\FSS} and optimally doped {\BFCA} and {\BFAP} can be unified within the aforementioned theoretical framework incorporating in the free energy the interplay between electronic nematicity, and $s$ and $d$ wave pairing instabilities --- $\varphi \Delta_s \Delta_d \cos\theta$. In the presence of nematic order and fluctuations, it is found that the electronic nematicity leads to deviations of $T_c$ [solid lines in Fig. 5(g)] from its original values [dashed lines in Fig. 5(g)].
The sign and magnitude of the deviation is essentially determined by the competition between $s$ and $d$ wave instability driven by controlling parameter such as doping levels, or the competition between different types of spin fluctuations favoring distinct pairing symmetries.
As shown in Fig 5(h), in optimally doped iron pnictides and at a lower concentration of {\FSS} [marked by blue arrow in Fig. 5(g)], $\beta<0$ [blue curve in Fig. 5(h)] indicates their superconducting electron pairing is dominated by $s_{\pm}$ wave instability. In contrast, at a higher doping of {\FSS} [marked by green arrow in Fig. 5(g)], an enhancement of $T_c$ [green curve in Fig. 5(h)] can be attributed to the emergence of competing $d$ wave instability. The coefficient $\beta$ increase monotonically across the NQCP, indicating the system is evolving towards an $s+d$ wave state. At the degeneracy point between $s$ and $d$ wave pairing, it is predicted that $\Delta T_c$ is a nonanalytic linear function, expressed as $\Delta T_c(\varepsilon)=|\alpha|\varepsilon$, which could occur at even higher doping $x>0.21$ and requires validation in future studies.
Nonetheless, the current results for $\Delta T_c(\varepsilon)$ provide compelling evidence concerning the significance of both $s$-and $d$-wave pairing instability in {\FSS} and suggest that the system evolves from a predominant $s_{\pm}$ wave towards an $s+d$ wave. Our study confirms the effectiveness of nematicity as a probe of pairing symmetry in FeSCs \cite{Fernandes13}.

In summary, we have demonstrated that uniaxial strain serves as an effective probe to elucidate the evolution of pairing symmetry in {\FSS} across its nematic regime. Our measurements of strain-induced changes in $T_c$ reveal a transition from an $s_{\pm}$-wave pairing state in undoped FeSe towards a mixed $s+d$-wave pairing state at higher sulfur doping. This evolution underscores the intricate interplay between nematicity and superconductivity. These findings not only establish uniaxial strain as a valuable tool for exploring superconducting pairing mechanisms but also suggest that competing electronic instabilities significantly influence the superconducting ground state in this class of materials. Further spectroscopic investigations under strain are essential to fully uncover the nature of the pairing symmetry near the nematic quantum critical point and beyond.

\begin{acknowledgments}
We thank Professor Pengcheng Dai for the helpful discussions. The work is supported by the National Key Projects for Research and Development of China (Grant No. 2021YFA1400400), the National Natural Science Foundation of China (Grant Nos. 12174029, and 11922402), and the Fundamental Research Funds for the Central Universities (Grant No. 2243300003).
\end{acknowledgments}



\begin{thebibliography}{}
$\\$
$\S$ These two authors contributed equally to this work.
$\\$

\bibitem{Fernandes2014} R. M. Fernandes, A. V. Chubukov, and J. Schmalian, What drives nematic order in iron-based superconductors? Nat. Phys. {\bf 10}, 97-104 (2014).

\bibitem{Bohmer2022} A. E. B\"ohmer {\it et al.},  Nematicity and nematic fluctuations in iron-based superconductors. Nature Physics {\bf 18}, 24-30 (2022).

\bibitem{Fernandes2022} R. M. Fernandes {\it et al.}, Iron pnictides and chalcogenides: a new paradigm for superconductivity. Nature {\bf 601}, 39-44 (2022).

\bibitem{Bohmer2018} A. E. B\"ohmer, and A. Kreisel, Nematicity, magnetism and superconductivity in FeSe. J. Phys. Condens. Matter {\bf 30}, 023001 (2018).

\bibitem{Coldea2018} A. Coldea and M. D. Watson. The key ingredients of the electronic structure of FeSe. Annu. Rev. Condens. Matter Phys. {\bf 9}, 125-146 (2018).

\bibitem{Coldea2021} A. I. Coldea, Electronic nematic states tuned by isoelectronic substitution in bulk FeSe$_{1-x}$S$_x$, Front. Phys. {\bf 8}, 594500 (2021).

\bibitem{Hsu2008} F. C. Hsu {\it et al.}, Superconductivity in the PbO-type structure $\alpha$-FeSe. Proc. Natl Acad. Sci. USA {\bf 105}, 14262 (2008).

\bibitem{McQueen2009} T. M. McQueen {\it et al.}, Tetragonal-to-orthorhombic structural phase transition at 90K in the superconductor Fe$_{1.01}$Se. Phys. Rev. Lett. {\bf 103}, 057002 (2009).

\bibitem{Hosoi2016} S. Hosoi {\it et al.}, Nematic quantum critical point without magnetism in FeSe$_{1-x}$S$_x$ superconductors. Proc. Natl Acad. Sci. USA {\bf 113}, 8139-8143 (2016).

\bibitem{Liu2024} Ruixian Liu {\it et al.}, Nematic Spin Correlations Pervading the Phase Diagram of FeSe${1-x}$S$_x$. Phys. Rev. Lett. {\bf 132}, 016501 (2024)

\bibitem{Licciardello2019} S. Licciardello {\it et al.}, Electrical resistivity across a nematic quantum critical point. Nature {\bf 567}, 213-217 (2019).

\bibitem{Dai_RMP} P. Dai, Antiferromagnetic order and spin dynamics in iron-based superconductors, Reviews of Modern Physics {\bf 87}, 855-896 (2015).

\bibitem{Chen19} T. Chen {\it et al.}, Anisotropic spin fluctuations in detwinned FeSe. Nat. Mater. {\bf 18}, 709-716 (2019).


\bibitem{Wang2016} Q. Wang {\it et al.}, Magnetic ground state of FeSe. Nat. Commun. {\bf 7}, 12182 (2016).

\bibitem{Lu2022} X. Lu {\it et al.}, Spin-excitation anisotropy in the nematic state of detwinned FeSe. Nat. Phys. {\bf 18}, 806-812 (2022).

\bibitem{Liu2024_2} R. Liu {\it et al.}, Nematic quantum disordered state in FeSe. arXiv:2401.05092.

\bibitem{Fernandes2018} J. Kang, A. V. Chubukov, and R. M. Fernandes, Time-reversal symmetry-breaking nematic superconductivity in FeSe. Phys. Rev. B {\bf 98}, 064508 (2018).

\bibitem{Sprau2017} P. O. Sprau {\it et al.}, Discovery of orbital-selective Cooper pairing in FeSe. Science {\bf 357}, 75-80 (2017).

\bibitem{Liu2018} D. F. Liu {\it et al.}, Orbital origin of extremely anisotropic superconducting gap in nematic phase of FeSe superconductor. Phys. Rev. X {\bf 8}, 031033 (2018).

\bibitem{Hashimoto2018} T. Hashimoto {\it et al.}, Superconducting gap anisotropy sensitive to nematic domains in FeSe. Nat. Commun. {\bf 9}, 282 (2018).

\bibitem{Sato2018} Y. Sato et al., Abrupt change of the superconducting gap structure at the nematic critical point in {\FSS}. Proc. Natl Acad. Sci. USA {\bf 115}, 1227-1231 (2018).

\bibitem{Hanaguri2018} T. Hanaguri {\it et al.}, Two distinct superconducting pairing states divided by the nematic end point in {\FSS}. Sci. Adv. {\bf 4} eaar6419 (2018)

\bibitem{Matsuura2023} K. Matsuura {\it et al.}, Two superconducting states with broken time-reversal symmetry in {\FSS}. Proc. Natl. Acad. Sci. USA {\bf 120} e2208276120 (2023)





\bibitem{Fernandes13} R. M. Fernandes {\it et al.}, Nematicity as a Probe of Superconducting Pairing in Iron-Based Superconductors. Phys. Rev. Lett. {\bf 111}, 127001 (2013).

\bibitem{Fisher2024} S. Ghosh {\it et al.}, Elastocaloric evidence for a multicomponent superconductor stabilized within the nematic state in Ba(Fe$_{1-x}$Co$_x$)$_2$As$_2$, arXiv:2402.17945.

\bibitem{Bartlett2021} J. M. Bartlett {\it et al.}, Relationship between Transport Anisotropy and Nematicity in FeSe, Phys. Rev. X {\bf 11}, 021038 (2021).

\bibitem{Yang2023} X. Yang {\it et al.}, In-plane uniaxial-strain tuning of superconductivity and charge-density wave in CsV$_3$Sb$_5$, Chin. Phys. B {\bf 32}, 127101 (2023).

\bibitem{SI} See Supplemental Material for details.

\bibitem{Mo2024} Z. Mo {\it et al.}, Cryogenic Digital Image Correlation as a Probe of Strain in Iron-Based Superconductors. Chin. Phys. Lett. 2024
DOI 10.1088/0256-307X/41/10/107102

\bibitem{Chu2012} J.-H. Chu, H.-H. Kuo, J. G. Analytis, and I. R. Fisher.
Divergent Nematic Susceptibility in an Iron Arsenide
Superconductor. Science {\bf 337}, 710-712 (2012).

\bibitem{Kuo2016} H.-H. Kuo {\it et al.}, Ubiquitous signatures of nematic quantum criticality in optimally doped Fe-based superconductors. Science {\bf 352}, 958-962 (2016).

\bibitem{Fisher2021} T. Worasaran, M. S. Ikeda, J. C. Palmstrom, J. A. W. Straquadine, S. A. Kivelson, I. R. Fisher, Nematic quantum criticality in an Fe-based superconductor revealed by strain-tuning. Science {\bf 372}, 973-977 (2021).

\bibitem{Ghini21} M. Ghini {\it et al.}, Strain tuning of nematicity and superconductivity in single crystals of FeSe. Phys. Rev. B {\bf 103}, 205139 (2021).

\bibitem{Malinowski2020} P. Malinowski {\it et al.}, Suppression of superconductivity by anisotropic strain near a nematic quantum critical point, Nat. Phys. {\bf 16}, 1189-1193 (2020).

\bibitem{Zhao2023} Z. Zhao {\it et al.}, Uniaxial-Strain Tuning of the Intertwined Orders in {\BFAP}. arXiv: 2305.04424.

\bibitem{ftnote} It is worth noting that $\Delta T_c(\varepsilon_{A_{1g}})$ is significant in the underdoped, nematic ordering regime of iron pnictides such as {\BFCA} and {\BFAP}.

\end{thebibliography}
\end{document}